\documentstyle[epsf,epsfig,preprint,aps,amssymb]{revtex}
\draft \preprint{SNUTP 02/015}
\begin{document}
\title{\Large\bf Orbifolded
$SU(7)$ and unification of families}
\author{
Kyuwan Hwang\footnote{kwhwang@phya.snu.ac.kr} and Jihn E.
Kim\footnote{jekim@phyp.snu.ac.kr}}
\address{
School of Physics, Seoul National University, Seoul 151-747,
Korea} \maketitle

\begin{abstract}
A 5D $SU(7)$ family unification model with two spinor
representations of $SO(14)$ is presented. The fifth dimension is
compactified on $S^1/Z_2\times Z_2'$. The orbifolding is used to
obtain 4D $SO(10)$ chiral fermions. The 4D grand unification group
is the flipped $SU(5)\times U(1)$. The doublet-triplet splitting
through the missing partner mechanism is achieved. Also, fermion
mass matrices are considered.
\\
\vskip 0.5cm\noindent [Key words: family unification, orbifold,
SU(7) grand unification, mass matrix]
\end{abstract}


\newpage
\newcommand{\bea}{\begin{eqnarray}}
\newcommand{\eea}{\end{eqnarray}}
\def\beq{\begin{equation}}
\def\eeq{\end{equation}}

\def\one{\bf 1}
\def\two{\bf 2}
\def\five{\bf 5}
\def\ten{\bf 10}
\def\tenb{\overline{\bf 10}}
\def\fiveb{\overline{\bf 5}}
\def\threeb{{\bf\overline{3}}}
\def\three{{\bf 3}}
\def\fb{{\overline{F}\,}}
\def\hb{{\overline{h}}}
\def\Hb{{\overline{H}\,}}

\def\slash#1{#1\!\!\!\!\!\!/}

The idea of grand unified theories(GUT's) is probably the most
influential one in particle physics in the last three
decades\cite{su5}. It was so attractive that some obstacles in
simple GUT models are expected to be resolved in a more complete
theory. One of the problems is the proton decay problem. In the
$SU(5)$ model, the proton lifetime is predicted to be of order
$M_{GUT}^4$ in units of GeV. The current experimental upper bound
on the partial decay rate into the $e^+\pi^0$ decay mode is $(1.6
\times 10^{33}\ {\rm yr})^{-1}$, which implies a huge $M_{GUT}>
10^{15}$~GeV. It is consistent with the significant separation of
the coupling constants of the strong, weak, and electromagnetic
interactions. This was considered as one of the successes of
GUT's. But this huge mass $M_{GUT}$ led to the so-called gauge
hierarchy problem, which in turn led to the developments of
technicolor, supersymmetry, and superstring in the last two
decades. Another problem in this huge $M_{GUT}$ is the
doublet-triplet splitting  problem in the quintet(${\bf 5}_H$)
Higgs that the standard model doublet Higgs boson is light($\sim
100$~GeV) while the accompanying color triplet boson is needed to
be supermassive. In most GUT models, one needs a fine-tuning to
achieve this doublet-triplet splitting.

Because of the dramatic success of GUT's in the unification of
coupling constants, the flavor problem(or the family problem),
which is the most important problem in the standard model, has
been expected to be resolved with the GUT idea\cite{georgi}. Let
us call this kind of unification the grand unification of
families(GUF). There have been attempts toward flavor unification
in larger GUT groups such as $SU(7)$ GUF\cite{kimsu7}, $SU(8)$ GUF
\cite{frampton}, etc., but the predictions given in any of these
models have not been confirmed. Therefore, it is fair to say that
the GUF attempts along this line has not led to any convincing
theory so far. On the other hand, in the heterotic superstring
models the representation {\bf 248} of $E_8$ is so large that the
known three families are believed to be contained in {\bf 248}.
Indeed, the superstring compactifications led to
phenomenologically interesting multi generation models
\cite{candelas,dixon,iknq}. In particular, the $Z_3$ orbifold
compactification has been very attractive since they give the
family number as multiples of 3. Also, it has been noted that the
doublet-triplet splitting problem is resolved in some orbifold
compactificaions\cite{iknq}.

The orbifold compactification is one of the efficient and simple
way to break down the huge heterotic string group $E_8\times
E_8^\prime$\cite{dixon}. However, the ten dimensional(10D)
superstring world is too far separated away from our low energy
four dimensional(4D) world. Therefore, the field theoretic
orbifold compactification(FTOC)\cite{kawamura} in five
dimension(5D) has attracted a great deal of attention recently
because of its simplicity, requiring only the field theoretic
information. In a sense, the FTOC is a bottom-up approach. In this
paper, we consider the FTOC even though a more fundamental theory
is based on the string theoretic orbifold
compactification(STOC)\cite{dixon}.

The initiation of FTOC started from the observation that the
doublet-triplet splitting can be understood by making the color
triplet boson superheavy, while the doublet Higgs boson can be
made a Kaluza-Klein(KK) zero mode by appropriately choosing the
charges of the discrete group in consideration. As noted in STOC,
the orbifold is known to have the mechanisms both for the
doublet-triplet splitting\cite{iknq} and for the unification of
flavor\cite{dixon,iknq}. In this regard, it is not unreasonable to
attempt the flavor unification also in FTOC as first tried in
\cite{barr}.

Along this FTOC line, we attempt to understand the flavor problem
in a 5D extended GUT, compactified on the orbifold $S^1/Z_2\times
Z'_2$\cite{difference}. The group $SU(6)$ cannot unify the flavor
since {\bf 15} of $SU(6)$ contains only one {\bf 10} of $SU(5)$.
The simplest GUT unifying the flavor is $SU(7)$. The $SU(7)$ model
of Ref.\cite{kimsu7} contains two standard families and two
non-standard families\cite{righthanded} among which one lepton
family becomes standard, but the others are unfamiliar ones. Alas,
due to the orbifolding in 5D instead of twisting the group, all
the unfamiliar families can be made familiar ones which can be
removed or kept depending on the $Z_2'$ charge. We note that the
${\ten}\oplus {\fiveb}$ of $SU(5)$\cite{su5} and ${\bf 35}\oplus
\overline{\bf 21}\oplus {\bf 7}$ of $SU(7)$\cite{kimsu7} models
are basically the $SO(10)$ and $SO(14)$ models with the spinor
representations for fermions, breaking down to $SU(5)$ and
$SU(7)$, respectively. Thus, the family unification hints toward
the chain $SU(2n+1)$ or $SO(4n+2)$. In this paper, we choose the
simplest generalization and {\it construct a GUF model in 5D
$SU(7)$ gauge group with the spinor representation(s) as the
matter assignment.} In this paper, $SO(14)$ is considered
interchangeably with $SU(7)$ up to a singlet\cite{kimsu7},
\begin{equation}  \label{spinor}
{\bf 64}=\psi^{ABC}+\psi_{AB}+\psi^A+{\bf 1}
\end{equation}
where the multi-indices imply the antisymmetric combinations, and
$A=1,2,\cdots,7$. When we say an $SU(7)$ spinor, it is meant
Eq.~(\ref{spinor}) without the singlet.

\bigskip

\noindent {\it Orbifold compactification}:  In 5D, the fifth
dimension $y=Rx_5$ is compactified on the circle $S^1$: $x_5
\equiv x_5 + 2\pi$. Points on $S_1$ are identified under the
$Z_2(x_5 \to -x_5)$ and $Z'_2(x_5 \to \pi - x_5)$. Let
 any fermion in $SU(7)$ tensor representation
has the following parity symmetry,
\begin{eqnarray}
\label{z2p} Z_2:\quad && \psi^{AB\cdots}(-x_5) =
\lambda_\psi \gamma_5
    P^A_{A'} P^B_{B'} \cdots \psi^{A'B'\cdots}(x_5),
\quad\quad
P \equiv {\rm diag} (I_5, I_2),  \\
Z_2':\quad && \psi^{AB\cdots}(\pi-x_5) = \lambda'_\psi \gamma_5
    P'^A_{A'} P'^B_{B'} \cdots
  \psi^{A'B'\cdots}(x_5),
\quad\quad P' \equiv {\rm diag} (I_5, -I_2),
\end{eqnarray}
where $I_n$ is the $n$ dimensional identity matrix, and $\lambda$
and $\lambda'$ are either $+1$ or $-1$. Due to the non-commuting
boundary conditions given by $P'$ in the group space, the gauge
group breaks down to
\begin{equation}
\label{break} SU(7) \longrightarrow SU(5)\times SU(2)_F\times
U(1) \,,
\end{equation}
where $SU(2)_F$ plays the role of family symmetry. Because of the
$SU(2)_F$, we expect that light two generations and the third heavy
generation are discriminated.

Since we start with a group containing $SU(5)$, there exists a
possibility that $U(1)$-electromagnetism contains an $SU(5)$
singlet piece\cite{su51} which is called the flipped $SU(5)$. The
flipped $SU(5)$ was extensively studied in fermionic construction
of 4D string models\cite{ellis}. The merit of the flipped $SU(5)$
in string models is that one does not need an adjoint
representation of $SU(5)$ for breaking $SU(5)$ down to the
standard model(SM). The $\psi^{\alpha\beta}({\ten})$ has a
$Q_{em}=0$ element $\psi^{67}=\nu^c$ which can have a GUT scale
vacuum expectation value(VEV), hence breaks the unified group to
the SM. At the same time, this VEV gives a large mass to the color
triplet Higgs fields through the missing partner mechanism as
discussed below\cite{missing}. Note that orbifolding is not needed
for the doublet-triplet splitting.

Therefore, let us choose the matter representation and the
$Z_2^\prime$ parity assignment $\lambda^\prime$ so that
$SU(5)\times U(1)$(the flipped $SU(5)$) is the GUT group. Under
this choice of $Z_2^\prime$ eigenvalues, the resulting zero modes
automatically form an anomaly free combination of $SO(10)$
spinors. The 4D chiral anomaly depends not only on the bulk matter
but also on the $Z_2^\prime$ parity assignment\cite{anomaly}.
However, our selection of $Z_2^\prime$ parity will give no anomaly
since the zero mode fermions form $SO(10)$ spinors.
  This property may be understood better if we consider the connection between
the two symmetry breaking chains
\begin{equation}  \label{so14}
SO(14)
\begin{array}{lll}
\raisebox{-3mm}{\vector(3,1){30}} & SO(10)\times SU(2)_F\times
SU(2)' &
\raisebox{1mm}{\vector(3,-1){30}} \\
\raisebox{6mm}{\vector(3,-1){30}}
& SU(7)\times U(1)' &
\raisebox{3mm}{\vector(3,1){30}}
\end{array}
SU(5)\times SU(2)_F \times U(1) \times U(1)'
\end{equation}

\bigskip

\noindent{\it Matter content}: A spinor of $SO(14)$ under the
breaking chain of Eq.(\ref{so14}) is
\begin{equation} \label{bulk}
\Psi^{ABC} \oplus \Psi_{AB} \oplus \Psi^{A} \oplus \Psi= {\bf
16}\otimes {\two}_F \oplus \overline{\bf 16} \otimes {\two}'\,,
\end{equation}
where the RHS is the decomposition into $SO(10)\times SU(2)\times
SU(2)'$ and the anti-symmetrization of the indices are assumed.
Since we are dealing with $SO(4n+2)$ groups, the models considered
do not have the anomaly problem.

A 5D $SO(14)$ spinor has four left-handed and four right-handed 4D
$SO(10)$ spinors. Under the torus compacification, these eight
$SO(10)$ spinors form four massive Dirac spinors and are removed
from the low energy spectrum. But twisting can allow some zero
modes. Let the $Z_2$ action in Eq.(\ref{z2p}) makes the
right-handed component of a 5D spinor heavy (breaking one
supersymmetry if there was). In other words, only 4 left-handed
$SO(10)$ spinors(one left-handed $SU(7)$ spinor) in 4D remain as
zero modes. It is represented under $SU(5)\times SU(2)\times U(1)$
as:
\begin{equation}
\begin{array}{llllllllll}
\Psi^{ABC} &=& \psi^{\alpha\beta\gamma} & {(\tenb,\one)}_6 &
\oplus
    & \psi^{\alpha\beta i} & {(\ten,\two)}_{-1}    & \oplus
    & \psi^{\alpha ij} & {(\five,\one)}_{-8}    \\
\Psi_{AB} &=& \psi_{\alpha\beta} & {(\tenb,\one)}_{-4} & \oplus
    & \psi_{\alpha i} & {(\fiveb, \two)}_3 & \oplus
    & \psi_{ij} & {(\one,\one)}_{10} \\
\Psi^{A} &=& \psi^{\alpha} & {(\five, \one)}_2 & \oplus
    & \psi^{i} & {(\one,\two)}_{-5} & & &
\end{array}
\end{equation}
where the total number of $\ten$ and $\tenb$ is four which is the
number of massless $SO(10)$ spinor zero modes. Here, the upper
case Roman letters $A,B,C,\cdots$ are the $SU(7)$
indices($1,2,\cdots,7$), the lower case Greek letters
$\alpha,\beta,\gamma, \cdots$ are the $SU(5)$
indices($3,4,\cdots,7$), and the lower case Roman letters $i,j$
are the $SU(2)_F$ indices 1, 2.
We can assign $\lambda' = -1$ to the $Z_2'$ parity  of
the whole $SU(7)$ spinor ($\Psi^{ABC}, \Psi_{AB}, \Psi^A$),
leaving the following zero modes
\begin{equation}
\label{zeromode} {(\ten,\two)}_{-1}\,,\quad
{(\fiveb,\two)}_{3}\,,\quad {(\one,\two)}_{-5}\,,
\end{equation}
which is exactly the anomaly free combination of the flipped
$SU(5)$ model\cite{su51}. Thus, this consistent choice of $Z_2'$
parity picks up one irreducible representation of ${\bf 16}\otimes
\two$ of $SO(10)\times SU(2)$ in 4D among the full spinor of
$SO(14)$ shown in Eq.(\ref{bulk}). The reason for this consistent
selection is in that a spinor of $SO(4n+2)$ can be decomposed into
the sum of alternating totally antisymmetric tensors of $SU(2n+1)$
as shown in Eq.~(\ref{bulk})\cite{su8}.

The 5D $SU(7)$ model presented above has two families, neatly
unified in a doublet of $SU(2)_F$ in Eq.~(\ref{zeromode}). We need
to introduce the third family. A simple choice is that the third
family is a singlet under $SU(2)_F$. We can put this $SU(2)_F$
singlet, ${(\ten,\one)}_{-1}\oplus{(\fiveb,
\one)}_{3}\oplus{(\one,\one)}_{-5}$ under $SU(5)\times U(1)$, at
the asymmetric fixed point. Then we need to put Higgs fields with
the gauge charges $ {\ten}_{-1},\quad {\tenb}_{1},\quad
{\five}_{2},\quad {\fiveb}_{-2}\, $ at the asymmetric fixed brain
also. $\ten$ and $\tenb$ are required to break $SU(5)\times U(1)
\to SU(3)_c\times SU(2)_L \times U(1)_Y$. $\five$ and $\fiveb$
contain the doublet Higgs for the $SU(2)_L\times U(1)_Y$ breaking
into $U(1)_{em}$.

In the remainder of this paper, however, we study a more
interesting case that the third family is also a member of an
$SU(2)_F$ doublet. In addition, let us extend to the {\it
supersymmetric case} so that the discussion on the Higgs
multiplets is neat. Put the same $SU(7)$ combination of
Eq.(\ref{bulk}) in the bulk again, from which we obtain the
additional zero modes given in Eq.~(\ref{zeromode}). Below the
$SU(2)_F$ breaking scale, one set of the $SU(2)_F$ doublet becomes
the third family fermions. The superpartners of the remaining
$SU(2)_F$ doublet can be Higgs multiplets: $H({\ten}_{-1})$,
$\overline{h}({\fiveb}_{3}), \phi({\one}_{-5})$\,. However,
$\overline h(\fiveb_{3})$ in the flipped $SU(5)$ does not have a
color triplet with $Q_{\rm em}=-1/3$; hence the $\threeb$
component of $H({\ten}_{-1})$ with $Q_{\rm em}=+1/3$ does not have
a partner in $\overline h({\fiveb}_{3})$, and the doublet-triplet
problem is not solved. To solve this doublet-triplet splitting
problem, we introduce ${\five}_2$ and ${\fiveb}_{-2}$ which have
color triplets with the needed electric charge. These may come
from $\bf 7\oplus \overline{7}$ of $SU(7)$, or $\bf 14$ of
$SO(14)$.

\bigskip

\noindent{\it Missing partner mechanism}: We introduced two
$SU(2)_F$--doublet spinors of $SU(7)$. For the Higgs fields, let
us introduce  ${\five}_2$ and ${\fiveb}_{-2}$ in the bulk, and in
addition \{${\tenb}_1\oplus {\five}_{-3} \oplus {\one}_5$\} at the
asymmetric fixed point, which are $SU(2)_F$--singlets. Toward a
detail discussion on the mass matrices of light fermions and the
doublet-triplet splitting mechanism, let us name two
$SU(2)_F$--doublets of $SO(14)$ spinor as
\begin{equation} \label{mdoub}
T_i({\ten}_{-1}), {\fb}_i({\fiveb}_3), E_i^c({\one}_{-5}), \quad
\mbox{\rm and} \quad T'_i({\ten}_{-1}), {\fb'}_i({\fiveb}_3),
E_i^{'c}({\one}_{-5}) ,
\end{equation}
where the family indices $i=1,2$ and $SU(2)_F$--singlets as
\begin{equation} \label{msing}
\Hb(\tenb_{1}), h'({\five}_{-3}), \phi({\one}_{5}), \quad
\mbox{\rm and} \quad h({\five}_2), \hb({\fiveb}_{-2})\,,
\end{equation}
and the components of each multiplet as
\begin{equation}
{\ten}_{-1} : \pmatrix{d^c & q \cr q & \nu^c} \quad\quad
{\fiveb}_{3} : \pmatrix{u^c \cr \ell} \quad\quad {\fiveb}_{-2} :
\pmatrix{\overline{D} \cr h^+ } \quad\quad {\five}_{+2} :
\pmatrix{D \cr h^- } \,
\end{equation}
where $\overline{D}$ and $h^+$ carries the hypercharge $1/3$ and
$1/2$, respectively.

In order to break the unified gauge group, we need two additional
$SU(2)_F$--doublet fields $\{\chi_i^1, \chi_i^2\}= 2{(\one,
\two)}_0$ at the asymmetric fixed point.
The superpotential relevant to the GUT symmetry breaking
and the masses of the third generation fermions,
written in the asymmetric fixed brain, are given by
\begin{equation} \label{wh}
W_{H}=\Hb\Hb\hb + T'T'h + T'\fb'\hb + \fb' E^{\prime c} h +
\fb'h'\chi^2 + E^{\prime c}\phi\chi^1
\end{equation}
This superpotential contains the most general cubic
terms of the singlet fields in Eq.(\ref{msing}) and
the primed doublet fields in Eq.(\ref{mdoub})
consistent with the following two discrete symmetries
\begin{equation} \label{disc}
Z_2^\chi\ :\ \chi^1\rightarrow -\chi^1\ ,\ \ \phi\rightarrow
-\phi, \qquad \quad Z_2^H\ :\ \Hb\rightarrow -\Hb
\end{equation}
while the other fields are invariant under $Z_2^\chi$ and $Z_2^H$.
We do not allow $h\hb$ term in the superpotential, which is
anticipated in the superstring models. By the development of VEV
along the $D$-flat(and $F$-flat) direction $(T' \Hb
\chi^1)(\chi^1\chi^2)$,
\begin{equation}
\langle\nu^c_{T'_1} \rangle = \langle\overline{\nu}^c_{\Hb}
\rangle = \frac{1}{\sqrt{2}}\langle\chi_2^1\rangle
=\langle\chi_1^2\rangle= M_{G},\label{symm}
\end{equation}
both $q_{T'_1}$ and $q_{\overline{H}}$ are either eaten by the heavy
gauge bosons or made heavy by the supersymmetric Higgs mechanism.
>From the superpotential terms in Eq.(\ref{wh})
the components $d^c_{T'_2}, D_h, \overline{d}^c_\Hb,
\overline{D}_\hb, \fb'_2,E_1^{\prime c}$ and $h'$ become massive
after the symmetry breaking, while
$h^+$ and $h^-$ remain massless and fulfil the doublet-triplet
splitting. The rest massless components \{$d^c_{T'_1}, q_{T'_2},
u^c_{F'_1}, \ell_{F'_1}, E^c_2$\} form the third generation family.

\bigskip

\noindent {\it Mass matrices}: In order to reproduce the realistic
fermion masses and mixing angles, we need an additional global
symmetry which prevents the light generation doublets $T,\fb, E^c$
from acquiring the same large mass as the third generation ones
$T', {\fb}', E^{c'}$. Here, as a simplest option available, we
just try an anomalous global $U(1)_F$ symmetry. Like the models
with $U(2)_F$ family symmetry in the literature \cite{u2}, if we
break the $SU(2)_F\times U(1)_F$ in two steps
\begin{equation}
\label{su2brk} SU(2)_F\times U(1)_F \stackrel{\epsilon
}{\longrightarrow} U(1) \stackrel{\epsilon'
}{\longrightarrow} \{e\}\,
\end{equation}
where $\epsilon \sim 0.02$ and $\epsilon' \sim 0.004$ in units of
a UV cutoff scale are the order parameters for each step, we can
suppress light generation masses by small parameters $\epsilon$
and $\epsilon'$. For a model consturction, let us assign $U(1)_F$
charge +1 to unprimed $SU(2)_F$--doublet fields, and 0 to the
other fields. In addition, let us introduce an $SU(2)_F$ singlet
$\phi(-1)$ and triplets $S_{\{ij\}}^{1,2}(-2)$($ij$ symmetric)
with the $U(1)_F$ charges indicated inside the parenthesis. The
relevant superpotential terms are given by,
\begin{eqnarray}
W_Y&=&\sum_{a=1,2}{1\over M_*}\left[S^{a} TTh + (\frac{\phi^2}{M_*}+S^a) T\fb \hb +
(\frac{\phi^2}{M_*}+S^a) \fb E^c h \right]\nonumber\\
&&+{\phi\over
M_*} \left[TT'h + (T\fb'+ T'\fb)\hb + (\fb E^{'c} + \fb' E^c)h
\right]\,\nonumber
\end{eqnarray}
where $M_*$ is the UV cutoff scale. Requiring the VEVs of the
$\lq$flavon' fields $\phi, S^{1,2}$ to be
\begin{equation}
\langle \phi \rangle \sim \epsilon M_*, \quad \langle
S_{\{22\}}^{1} \rangle \sim \epsilon M_*, \quad \langle
S_{\{12\}}^{2} \rangle \sim \epsilon' M_*\,,
\end{equation}
the mass matrices look like
\begin{equation}
{M^{u,d}\over M^{u,d}_{33}} \approx
  \pmatrix { 0 & \epsilon' & 0 \cr
 \epsilon' & \epsilon & \epsilon \cr \epsilon & 0 & 1}\,,
\qquad\quad
{M^{e}\over M^{e}_{33}} \approx
  \pmatrix { 0 & \epsilon' & \epsilon \cr
 \epsilon' & \epsilon & 0 \cr 0 & \epsilon & 1}\,.
\end{equation}
This form of mass matrices gives the qualitatively correct mass
spectrum and CKM mixing matrix elements. If we let the two
symmetry breaking steps in Eq.(\ref{su2brk}) occur with a single
triplet $S_{\{ij\}}$ instead of two different triplets
$S_{\{ij\}}^{1,2}$, the $SU(2)_F$ symmetry would enforce the
unrealistic relation $ m_u/m_c=m_d/m_s=m_e/m_\mu$ precisely, as
long as the mixing between light two generations and the third
generation remains small. In our model, however, the discrepancy between
$m_u/m_c, m_d/m_s$ and $m_e/m_\mu$ as well as $m_c/m_t, m_s/m_b$
and $m_\mu/m_\tau$ can be accounted for by the numerical
coefficients of tolerable size, since the up-type quark, down-type
quark and lepton masses come from different superpotential terms.

In this paper, we constructed a 5D $SU(7)$(or $SO(14)$) GUF model
with two spinors of $SO(14)$, with the orbifold compactification
$S_1/Z_2\times Z_2'$, which realizes the three families of
fermions in the flipped $SU(5)$ and the doublet-triplet splitting
of Higgs multiplet. We introduced ${\bf 5}_{+2}$ and
$\overline{\bf 5}_{-2}$, an $SO(10)$ vector arising from the
$SO(14)$ vector {\bf 14}. There may be a deep reason for the two
5D $SO(14)$ spinors. In the $E_8\times E_8'$ heterotic string
model, the adjoint or the fundamental representation of $E_8$,
$\bf 248$, contains $\bf 128 \oplus 120$ of $SO(16)$, one of the
maximal subgroup of $E_8$. The $SO(16)$ spinor {\bf 128}
decomposes to two $SO(14)$ spinors: $\bf 64 + \overline{64}$. For
$\overline{\bf 64}$({\bf 64}), we pick up the
right-handed(left-handed) components and hence assign $-$($+$ as
before) for the $Z_2$ quantum number $\lambda$ so that the
massless modes are the left-handed fields with the combination
given in (\ref{bulk}). This may be the reason that nature chooses
two $SO(14)$ spinors. Also, the anomalous $U(1)_X$ we introduced
to discriminate $\overline{\bf 64}$ from $\bf 64$ could come from
$E_8/SO(16)$, which assign chiral charge to $SO(14)$ spinors. On
the other hand, {\bf 120} of $SO(16)$ breaks down to ${\bf
91}\oplus$~two~${\bf 14}$'s~$\oplus {\bf 1}$ and the needed
$SO(14)$ vector {\bf 14} can be assigned to {\bf 120} of $SO(16)$.

\acknowledgments This work is supported in part by the BK21
program of Ministry of Education, the KOSEF Sundo Grant, and by
the Office of Research Affairs of Seoul National University.


\begin{thebibliography}{99}
\bibitem{difference}
Our motivation for
FTOC is different from that of Ref.\cite{kawamura} in that we want
to unify the families while Ref.\cite{kawamura} tries to split the
$SU(5)$ quintet Higgs boson.
\bibitem{righthanded} In Ref.\cite{kimsu7}, it was possible to have
chiral families due to the twisting of the electromagnetic charges
of two right-handed families.
\bibitem{su8} One might want to unify all 3 generations in $SU(8)$.
However, our mechanism does not work for $SU(2n)$ models, since an
$SO(4n)$ spinor always branches to a vectorlike representation of
$SU(2n)$.
\end{thebibliography}
\end{document}